\documentclass{svmult}
\usepackage{amsmath}
\usepackage{graphicx}
\usepackage[authoryear,round,longnamesfirst]{natbib}
\begin{document}
\title*{Dual Scaling of Rating Data}
\author{Michel van de Velden, Patrick J.F. Groenen}

\institute{Michel van de Velden \at Erasmus University Rotterdam, \email{vandevelden@ese.eur.nl}, Patrick J.F. Groenen \at Erasmus University Rotterdam, \email{groenen@ese.eur.nl}}
\maketitle

\abstract{}

When applied to contingency tables, dual scaling and correspondence are mathematically equivalent methods. For the analysis of rating data, however, the methods differ. To a large extent this is due to differences in pre-processing of the data. In particular, in dual scaling, ratings are either transformed to rank order, or to successive category data before applying a customised dual scaling approach. In correspondence analysis, on the other hand, a so-called doubling of the original ratings is applied before applying the usual correspondence analysis formulas. In this paper, we consider these differences in detail. We propose a dual scaling variant that can be applied directly to the ratings and we compare theoretical as well as practical properties of the different approaches. 

\section{Introduction}
The works of Nishisato has shown that dual scaling is a powerful method that can be applied to solve a wide variety of data analysis problems. Dual scaling is closely related and, for many practical purposes and applications, equivalent to correspondence analysis. The books of  \cite{Nishisato1994} and \cite{Greenacre1984} give a detailed account of the relationships and origins of the methods. Mathematically, relationships are particularly strong (see, e.g., \cite{Greenacre1984, Vandevelden2000phd}). Perhaps, the biggest difference between the methods concerns correspondence analysis' focus on geometry versus dual scaling's emphasis on the optimal scaling properties. 

Although both correspondence analysis and dual scaling are often considered for analysing a contingency table, both methods can be applied to other types of data. However, with respect to the analysis of such other data types, the approaches do in fact differ. %For the analysis of rank order \cite{Vandevelden2000} and  paired comparison data \cite{Vandevelden2004} some relationships and properties were studied. 
In this Chapter, we explicitly consider dual scaling and correspondence analysis of rating data. 

For correspondence analysis, the analysis of rating data is explicitly treated in \cite[][Chapter 6]{Greenacre1984}. However, \cite{Greenacre2017} treats the analysis of rating data in a chapter titled ``Data re-coding" (Chapter 23). The new labelling of the topic is a direct result of the way the analysis of rating data is defined in the correspondence analysis literature. That is, for the analysis of rating data, the rating data are first re-coded in a specific form and subsequently the usual correspondence analysis calculations are applied to the re-coded data. 

For dual scaling, the analysis of rating data is treated in the context of paired comparison and successive categories data. The proposed methods in these contexts also amount to the application of the usual dual scaling calculations (which are equivalent to the correspondence analysis calculations) to re-coded data. However, as we show in this paper, the re-coding in dual scaling and correspondence analysis is different and, consequently, properties of the solutions differ as well. Note that, a direct analysis of rating data does not appear to exist in the dual scaling literature. However, as we show in Section \ref{SectDSRating}, we can tackle this problem by using a similar interpretation of the ratings as done in the correspondence analysis literature.

In this paper, we review the existing approaches to the analysis of rating data in dual scaling and correspondence analysis. We do so by first briefly summarising the different types of re-coding in Sections \ref{SectCARating} and \ref{SectDSRating}. Furthermore, we propose a method that allows a more direct treatment of rating data in dual scaling.  Next, using the optimal scaling framework that is fundamental in the works of Nishisato, we provide insights into the theoretical differences between the methods, and we discuss the implications of these differences in practice. % when applying either duals caling or correspondence analysis of ratings. 
We illustrate the differences by means of an example data set taken from \cite{Nishisato1994} and provide some final remarks in Section \ref{SectFinal}

\section{Dual Scaling} \label{SectDS}
The objective of dual scaling is to find optimal scaling values or scores (or coordinates) for row categories that maximise the between row variance whilst at the same time finding scores for the column categories that maximise the between column variance. Here we only give the basic formulas needed to calculate the dual scaling solution for analysing a two-way data table $\mathbf{F}$ consisting of non-negative integers. For a complete description of the rationale and a derivation of the dual scaling solution see \cite{Nishisato1994}. 

Let $\mathbf{F}$ denote an $n \times p$ matrix consisting of non-negative entries and define diagonal matrices $\mathbf{D}_{r}$ and $\mathbf{D}_{c}$ in such a way that $\mathbf{D}_{r}\mathbf{1}_{p}= \mathbf{F}\mathbf{1}_{p} = \mathbf{r}$ and $\mathbf{D}_{c}\mathbf{1}_{n}= \mathbf{F}^{T}\mathbf{1}_{n} = \mathbf{c}$, where, generically, $\mathbf{1}_{i}$ denotes an $i \times 1$ vector of ones. Consider the singular value decomposition
\begin{equation}
    \mathbf{D}_{r}^{-1/2}\left( \mathbf{F} - \frac{1}{s} \mathbf{rc}^{T} \right)\mathbf{D}_{c}^{-1/2} = \mathbf{U}\mathbf{\Lambda}\mathbf{V}^{T}, 
\end{equation}
where $s=\mathbf{1}_{n}\mathbf{F}\mathbf{1}_{p}$, and, without loss of generality, the singular values on the diagonal of $\mathbf{\Lambda}$ are in non-increasing order. 
The $k$-dimensional optimal scaling values (i.e., the scores/coordinates) for rows and columns are $\mathbf{X}=\mathbf{D}_{r}^{-1/2}\mathbf{U}_{k}$ and $\mathbf{Y}=\mathbf{D}_{c}^{-1/2}\mathbf{V}_{k}$ respectively, where $\mathbf{U}_{k}$ and $\mathbf{V}_{k}$ correspond to the first $k$ columns of $\mathbf{U}$ and $\mathbf{V}$. 

Note that by defining $\mathbf{X}$ and $\mathbf{Y}$ in this way, they are standardised such that $\mathbf{X}^{T}\mathbf{D}_{r}\mathbf{X} = \mathbf{Y}^{T}\mathbf{D}_{c}\mathbf{Y} = \mathbf{I}_{k}$. In the correspondence analysis literature, the matrices $\mathbf{X}$ and $\mathbf{Y}$ are referred to as standard coordinates. Alternatively, defining $\mathbf{G}=\mathbf{D}_{r}^{-1/2}\mathbf{U}_{k}\mathbf{\Lambda}_{k}$ and $\mathbf{H}=\mathbf{D}_{c}^{-1/2}\mathbf{V}_{k}\mathbf{\Lambda}_{k}$ gives the solution in so-called principal coordinates. For more details on the different scalings and their implications on, in particular, graphical representations of results, see \cite{Greenacre2017}. 

\section{Correspondence Analysis of Ratings (CAr)} \label{SectCARating}
%Crucial in correspondence analysis of ratings is the so-called doubling of the ratings. That is, rather than considering only the original ratings for the objects, for each object the rating on the reversed scale is added to the data table. 
Let $\mathbf{R}$ denote an $n \times p$ matrix of ratings on a $1$ to $q$ scale. We use an artificial example of $n=4$ individuals who each rate $p = 3$ objects on a $q = 5$ point rating scale, to illustrate the different data pre-processing steps required in the different variant. That is, 
\begin{eqnarray} \label{R}
  \mathbf{R} = \begin{bmatrix}
  2 & 4 & 5 \\
  3 & 3 & 1 \\
  2 & 1 & 4 \\
  1 & 5 & 3 
  \end{bmatrix}.
\end{eqnarray}
Correspondence analysis is concerned with count data. The ratings can be considered as counts by considering a rating value as the number of times an object was preferred over the lowest rating number. To achieve this, we simply subtract $1$ from the originals ratings. Let $\mathbf{T} = \mathbf{R} - \mathbf{1}_{n}\mathbf{1}_{p}^{T}$, denote the resulting matrix with values from $0$ to $q-1$. That is, if the original rating scale consists of $q$ ratings, we first subtract $1$ which leads in our toy example to
\begin{eqnarray*}
  \mathbf{T} = \mathbf{R} - \mathbf{1}_{4}\mathbf{1}_{3}^{T} = \begin{bmatrix}
  2 & 4 & 5 \\
  3 & 3 & 1 \\
  2 & 1 & 4 \\
  1 & 5 & 3 
  \end{bmatrix}  - 
  \begin{bmatrix}
  1 & 1 & 1 \\
  1 & 1 & 1 \\
  1 & 1 & 1 \\
  1 & 1 & 1 
  \end{bmatrix} = 
  \begin{bmatrix}
  1 & 3 & 4 \\
  2 & 2 & 0 \\
  1 & 0 & 3 \\
  0 & 4 & 2 
  \end{bmatrix} .
\end{eqnarray*}
Thus, $\mathbf{T}$ can be interpreted as the number of scale points below a given rating, or, equivalently, as the number of times an object was considered to exceed a threshold on the original rating scale. 

Mathematically, we can apply correspondence analysis to the count data in $\mathbf{T}$. However, the problem with such a procedure is that the direction of the original rating scale influences the results; reversing the scale would lead to different results. That is, if data are gathered on a scale were the lowest rating ($1$) corresponds to ``bad" and the highest rating ($q$) to ``good", and we decide to switch the labelling from $1$=``good" to $q$=``bad", the results of the analysis would change. Clearly this sensitivity to the direction of the scale is an undesirable effect.

To overcome this problem, the data are ``doubled", meaning that the rating data for both directions of the rating scale are considered simultaneously. In correspondence analysis, this is done by, for each object, adding a column with the rating on the reversed scale. Consequently, instead of $p$ columns, we obtain a matrix consisting of $2p$ columns. Let $\mathbf{S}$ denote the matrix of ratings on the reversed scale, that is, $\mathbf{S}=(q-1)\mathbf{1}_{n}\mathbf{1}_{p}^{T}-\mathbf{T}$. In our running example, we get
\begin{eqnarray*}
  \mathbf{S}=(q-1)\mathbf{1}_{n}\mathbf{1}_{p}^{T}-\mathbf{T} =  
  \begin{bmatrix}
  4 & 4 & 4 \\
  4 & 4 & 4 \\
  4 & 4 & 4 \\
  4 & 4 & 4 
  \end{bmatrix} - 
  \begin{bmatrix}
  1 & 3 & 4 \\
  2 & 2 & 0 \\
  1 & 0 & 3 \\
  0 & 4 & 2 
  \end{bmatrix} = 
  \begin{bmatrix}
  3 & 1 & 0 \\
  2 & 2 & 4 \\
  3 & 4 & 1 \\
  4 & 0 & 2 
  \end{bmatrix}.
\end{eqnarray*}
We construct a column-wise doubled matrix as $\mathbf{F}_{c}=\left[{ \mathbf{T}~|~\mathbf{S}} \right]$:
\begin{eqnarray*}
  \mathbf{F}_{c}=\left[{\mathbf{T}~|~\mathbf{S} }\right] =  
  \left[\begin{array}{ccc|ccc}
  1 & 3 & 4~ & ~3 & 1 & 0\\
  2 & 2 & 0~ & ~2 & 2 & 4\\
  1 & 0 & 3~ & ~3 & 4 & 1 \\
  0 & 4 & 2~ & ~4 & 0 & 2
  \end{array}\right].
\end{eqnarray*}
Substituting this doubled matrix $\mathbf{F}_{c}$ for $\mathbf{F}$ in the formulas of Section \ref{SectDS} yields the correspondence analysis solution. 

The specific structure of the doubled matrix $\mathbf{F}_{c}$ results in structured coordinates for the columns as well. In particular, the points corresponding to the same object, with the reversed ratings, can be connected through a straight line running through the origin (however, the distances from the origin of both points differ). \cite{Greenacre2017} uses this relationship and shows that the resulting lines can be divided into $q-1$ equal sized intervals with the endpoints corresponding to the endpoints of the rating scale. That is, rating $q$ corresponds to the point corresponding to the original rating, and rating $1$ corresponds to the point on the reversed scale. The approximated average rating value (on the original scale) can then be inferred from this plot by considering the value on this line at the origin. Furthermore, similar to the case in principal component analysis, the angles (at the origin) between the lines corresponding to the different attributes, approximate correlations between the ratings for the attributes. In fact, as shown in \citet[][pp. 103-104]{Vandevelden2004}, the analysis of the doubled matrix $\mathbf{F}_{c}$ is equivalent to a principal component analysis of a particularly scaled and double centred version of the original rating data. 

\section{Dual Scaling of Rating Data} \label{SectDSRating}
Dual scaling of rating data is not treated as topic of its own in \cite{Nishisato1994}. Instead, in the context of paired comparison and rank order data, \cite{Nishisato1994} proposes two dual scaling variants that require different re-coding of the data. The first approach requires re-coding of the ratings as rankings while the second approach involves a joint ranking of objects and, unobserved, rating boundaries. To these two approaches, we add a third, more direct, re-coding that relies on an interpretation of ratings similar to the one used in correspondence analysis and described in Section \ref{SectCARating}. In the following subsections, we briefly discuss these three types of re-coding as well as the dual scaling analysis of them. For convenience, we have labelled these dual scaling variants DS1 up to DS3.
%pre-processing/transformations. 1) Converting ratings to rank order data. These approach require different data 1) dual scaling of successive categories transformations. 

\subsection{Converting Ratings to Rank Order Data (DS1)}
For the first variant, rather than considering the observed ratings directly, for an observation $i$, one counts the number of times that individual $i$'s rating for object $j$ is rated higher than ratings for all other objects. This is equivalent to transforming the ratings to ranked (from $0$ to $p-1$) data and requires a way to deal with ties (i.e., equal ratings). For the data from our running example, that is the matrix $\mathbf{R}$ given in (\ref{R}), we get
\begin{eqnarray*}
  \mathbf{T}^* = 
  \begin{bmatrix} 
  0 & 1 & 2 \\
  1.5 & 1.5 & 0 \\
  1 & 0 & 2 \\
  0 & 2 & 1 
  \end{bmatrix},
\end{eqnarray*}
and, on the reversed scale,
\begin{eqnarray*}
  \mathbf{S}^* = \begin{bmatrix} 
  2 & 1 & 0 \\
  .5 & .5 & 2 \\
  1 & 2 & 0 \\
  2 & 0 & 1 
  \end{bmatrix}.
\end{eqnarray*}
Since the focus is now on the rank of the three objects, not their rating on a 5-point scale, this method clearly incurs a loss in information as only the direction of the difference is considered, and not the magnitude. 

To analyse the resulting rank order data \cite{Nishisato1994} proposes to construct a dominance matrix $\mathbf{E}$ consisting of the difference between the number of times an object was preferred over the other objects ($\mathbf{T}^*$) and the number of times it was not preferred over other object ($\mathbf{S}^*$). For our example we get
\begin{eqnarray*}
  \mathbf{E}=\mathbf{T}^*-\mathbf{S}^*=  
    \left[\begin{array}{rrr}
  -2 &  0 &  2 \\
   1 &  1 & -2 \\
  0 & -2 &  2 \\
  -2 & 1 &  1 \end{array}\right] .
  \end{eqnarray*}
  So that the sum of each row is zero. Note that the dominance matrix $\mathbf{E}$ contains positive and negative values. Moreover, as the row sums are all zero the usual dual scaling calculations, as set out in Section \ref{SectDS}, cannot be applied directly. \cite{Nishisato1994} resolves this by defining $\mathbf{D}_{r}=p(p-1)\mathbf{I}_n$, and $\mathbf{D}_{c}=n(p-1)\mathbf{I}_p$ respectively. Alternatively, as shown by \cite{Vandevelden2000}, one can apply the usual dual scaling approach to the row-wise doubled matrix $\mathbf{F}_{r}= \left[\mathbf{T}^{*^T}~|~ \mathbf{S}^{*^T} \right]^{T}$ yielding, for our example data,
\begin{eqnarray*}
  \mathbf{F}_{r}=   
  \left[\begin{array}{rrr}
  0 & 1 & 2 \\
  1.5 & 1.5 & 0 \\
  1 & 0 & 2 \\
  0 & 2 & 1 \\ \hline
  2 & 1 & 0 \\
  .5 & .5 & 2 \\
  1 & 2 & 0 \\
  2 & 0 & 1 
  \end{array}\right].
\end{eqnarray*}
Analysing the row-wise doubled matrix yields $2n$ scores for the $n$ rows. The scores for the first $n$ rows corresponds to the observations (rankings) on the original scale while the scores for the second set correspond to the observations (rankings) on the reversed scale. These two sets of scores, however, are trivially related as the scores in the second set are simply $-1$ times those in the first set. 

\subsection{Converting Rating Data to Successive Category Data (DS2)}
The second approach, introduced in \cite{Nishisato1980} and further developed in \cite{NishisatoSheu1984}, requires the introduction of ``boundaries", marking the difference between rating scale values. To each boundary we assign a rating that lies between the two values of the rating scale that the boundary represents. The observed ratings in $\mathbf{R}$ and the boundaries in $\mathbf{R}_{\text{bound}}$ are jointly ranked, resulting in so-called successive category data $\mathbf{R}_{\text{SCD}}$. Note that, in this way, in addition to the $p$ objects, $q-1$ boundaries are added as columns to the data matrix. In our example, using $1.5$ up to $4.5$ as ``rating" values for the boundaries, we get
\begin{eqnarray*}
  [\mathbf{R}~|~\mathbf{R}_{\text{bound}}] = 
  \left[\begin{array}{rrr|rrrr}
  2 & 4 & 5 & 1.5 & 2.5 & 3.5 & 4.5 \\
  3 & 3 & 1 & 1.5 & 2.5 & 3.5 & 4.5 \\
  2 & 1 & 4 & 1.5 & 2.5 & 3.5 & 4.5 \\
  1 & 5 & 3 & 1.5 & 2.5 & 3.5 & 4.5 
  \end{array}\right] ~~\rightarrow ~~~\mathbf{R}_{\text{SCD}} = 
  \left[\begin{array}{rrrrrrr}
  2 & 5 & 7 & 1   & 3   &  4  & 6   \\
  4.5  & 4.5  & 1 &  2  &  3  &  6  &  7 \\
  3 & 1 & 6 & 2 &  4  &   5  &  7  \\
  1 & 7 & 4 &  2  &  3  &  5  &  6  \\
  \end{array}\right].
\end{eqnarray*}
The resulting $n \times (p+q-1)$ matrix of rank ordered data $\mathbf{R}_{\text{SCD}}$ can be analysed in the same way as described above, that is, using the row-wise doubled matrix. Note that it doesn't matter what the exact values are that we insert for the boundaries, as long as they are between the actual ratings. 

As the boundaries are always ordered in the same way, the one-dimensional dual scaling solution for successive category data typically seems appropriate and sufficient in terms of explained variance. Moreover, as the successive category values for an individual are based on all ratings by the same individual, individual specific scale use is taken into account. For this reason, this specific coding was used by \cite{Schooneesetal2015} and \cite{Takagishietal2019} to construct methods to study response style bias in questionnaires. %In this paper, we do not further consider this approach. 

\subsection{Converting Rating Data to Count Data (DS3)} \label{SubsectDS}
%The approach converting the ratings to rank order data completely ignores the magnitude of the differences in ratings. In the successive category approach, the sizes of the differences are accounted for. However, the inclusion of the boundaries, which are by definition always ordered in the same way, a dominant first solution, appears. 
Both re-coding methods described in the previous subsections yield individual specific rankings. Consequently, the transformed rating values are individual specific as well. This implies that if individual $i$ assigns rating $j$ to an object, and individual $l$ assigns the same rating value $j$ to that object, the re-coded values do not have to be the same for both observations. If the actual ratings are considered to be meaningful and non-individual specific, this may not be a desirable property. To overcome this problem, one can re-code and interpret the ratings as previously described in Section \ref{SectCARating} in the context of correspondence analysis, that is, the rating values are re-coded to $\mathbf{T}$; the number of times an observation exceeds the boundaries on the original rating scale.

As before, we cannot apply dual scaling directly to $\mathbf{T}$ as this would lead to results that depend on the direction of the scale. Following the dual scaling approach for rank order data, we can overcome this by constructing a row-wise doubled matrix $\mathbf{F}_{r}=  \left[\mathbf{T}^T~|~ \mathbf{S}^T \right]^{T}$ and applying dual scaling to this matrix. From here on, we refer to this DS3 approach as dual scaling of rating data.

\section{Optimal Scaling Properties}% of dual scaling and correspondence analysis of rating data}

As seen in Sections \ref{SectCARating} and \ref{SectDSRating}, the difference between dual scaling and correspondence analysis of rating data amounts to a difference in doubling of the observed ratings after converting them to a $0$ to $q-1$ scale. That is, correspondence analysis of ratings is defined as correspondence analysis of $\mathbf{F}_{c}$ whereas dual scaling of ratings is defined as dual scaling of $\mathbf{F}_{r}$. To better understand the implications of these differences, we briefly review the optimal scaling properties of them. 

As shown by \cite{Nishisato1978} and \cite{Vandevelden2004}, the object scores obtained in an analysis of the dominance matrix $\mathbf{E}$, and, hence, the object scores in the analysis of $\mathbf{F}_{r}$, are equivalent to the optimal scaling values as defined and derived by \cite{Guttman1946}. As such, these values are determined ``so as to best distinguish between those things judged higher and those judged lower for each individual", \cite{Guttman1946}. Both \cite{Guttman1946} and \cite{Nishisato1978} explicitly consider paired comparison data. However, crucial in the formulation of the optimal scaling framework are the matrices $\mathbf{T}$ and $\mathbf{S}$. Hence, using these matrices in the context of rating data where, respectively, the entries represent the times an object rating exceeds or does not exceed the available rating boundaries, the optimal scaling properties remain valid.  That is, in dual scaling of ratings as defined in Section \ref{SubsectDS}, the scale values for the objects are assigned in such a way that they best distinguish between objects. 

In \cite{Guttman1946}, an optimal scaling solution for the individuals is not considered. However, we can rephrase \cite{Guttman1946}'s optimal scaling goal towards finding scale values for the observations/individuals as follows: Find scale values for individuals so as to best distinguish between individuals that judged an object higher and lower, for each object. Where once again higher (lower) indicates how often a rating exceeded (did not exceed) the boundaries. This ``dual" problem was considered, in the context of paired comparison data, by \cite{Vandevelden2004} who showed that the resulting optimal scaling values for individuals can be obtained by applying dual scaling/correspondence analysis to the column-wise doubled matrix $\mathbf{F}_{c}$. As before, interpreting the entries of $\mathbf{T}$ and $\mathbf{S}$ as the times an object rating exceeds or does not exceed the available rating boundaries, these optimal scaling properties remain valid in the analysis of $\mathbf{F}_{c}$. Hence, whereas dual scaling of ratings yields optimal scaling values for the objects, correspondence analysis of ratings yields optimal scaling values for the individuals. 

%An implication from the previous results is that, when defining optimal scaling values according to the framework and rationale as presented by Guttman \cite{Guttman1946}, the typical duality associated with a dual scaling (and correspondence analysis) solution obtained using the formulas of Section \ref{SectDS}, does not immediately carry over when we have rating data. That is, Guttman's optimal scaling values for subjects and objects based on rating data cannot be obtained simultaneously. 

Note that for the dual scaling analysis of rating data, the optimal scaling values for the doubled rows (i.e., individuals' ratings according to the original and reversed scales) are optimal in the usual dual scaling sense. That is, they maximize the variation between the rows of the doubled table. Similarly, for the correspondence analysis solution, the values for the doubled columns are optimal scaling values (i.e., the objects rated according to the original and reversed scales). However, when defining optimal scaling values according to the framework and rationale as presented by Guttman \cite{Guttman1946}, the typical duality associated with a dual scaling (and correspondence analysis) solution obtained using the formulas of Section \ref{SectDS}, does not immediately carry over when we have rating data. That is, Guttman's optimal scaling values for individuals and objects based on rating data cannot be obtained simultaneously. 

We summarised some properties of the different variants in Table \ref{Table:summary}. 
\begin{table}[]
\centering
\begin{tabular}{c|l|l|l}
    Method & Intervals & Optimal scaling & Doubling\\ \hline
    DS1 & Rank order only & Objects & Implicit; row-wise \\    DS2 & Successive categories & Objects and boundaries & Implicit: row-wise\\
    DS3 & Differences between ratings & Objects & Explicit: row-wise\\
    CAr & Differences between ratings & Individuals & Explicit: column-wise\\
\end{tabular}
\caption{Properties of the different variants for the analysis of rating data}
\label{Table:summary}
\end{table}

In summary, the difference between the dual scaling and correspondence analysis of rating data approaches amounts to a different way of dealing with the direction of the rating scales. For the dual scaling of rating data, as introduced in Section \ref{SubsectDS}, a row-wise doubling is employed. For the correspondence analysis of rating data, a column-wise doubling is used to resolve the problem. The effect of these different data pre-processing steps is that in the dual scaling analysis of rating data, the values for the objects are optimal scaling values whereas in the correspondence analysis of rating data, the coordinates for the individuals are optimal scaling values. 

In order to choose one method over the other, it is important to understand these differences. Depending on the type of application and the specific research goals, a choice can be made. Dual scaling of rating data may be more appropriate when one's prime concern is a visualisation (or quantification) of a set of objects based on the observed differences in the ratings for these objects. This could be the case, when, for example, relative positions of products based on how they are perceived by a group of individuals. One the other hand, if one is more concerned with a visualisation (or quantification) of the individuals, based on differences in rating patterns for a set of objects, correspondence analysis of rating data may be better equipped to pick up on the individual differences. 

\section{Applications}
To illustrate the differences between the dual scaling and correspondence analysis of rating approaches, we analyse an example data set from \cite[][p.230]{Nishisato1994} on perceived seriousness of crimes. In particular, we focus on the effects of the different doublings; that is, the analysis of a column-wise (CAr) and row-wise (DS3) doubled matrix of ratings. 

A sample of $17$ individuals indicated, on a rating scale from $1$ (``somewhat serious") to $4$ (``extremely serious"), the perceived seriousness of the following $8$ types of crimes: Arson, burglary, counterfeiting, forgery, homicide, kidnapping, mugging and receiving stolen goods. For ease of reproducibility, we included the data here in Table \ref{tab:DataTable}. Note that all individuals considered ``homicide" to be extremely serious. For this lack in variation, which leads to a singular $\mathbf{D}_c$ matrix in CAr, we removed this type of crime from our analyses. In the DS3 approach such a singularity does not occur and \cite{Nishisato1994} analyses the data without removal of this object.

\begin{figure}[ht]
    %\centering
    \begin{center}
\includegraphics[scale=.3]{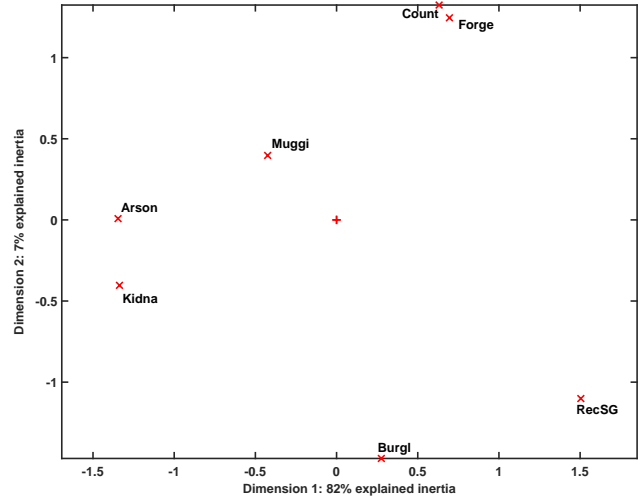}
    \caption{Dual scaling of ratings (DS3) for the crime perception data. Optimal scaling values for objects (crimes) in standard coordinates}\label{fig:FigDS_crimes}
\end{center}
    
\end{figure}

\begin{figure}[ht]
    %\centering
    \begin{center}
\includegraphics[scale=.3]{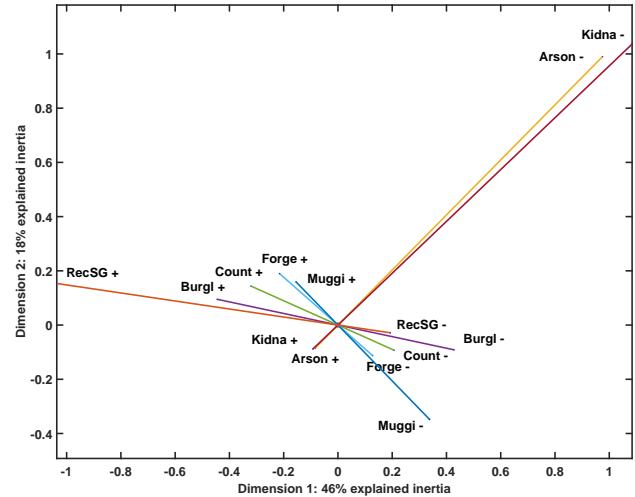}
    \caption{Correspondence analysis of ratings (CAr) for the crime perception data. Objects (crimes) in principal coordinates}
    \label{fig:FigCA_crimes}
\end{center}
    
\end{figure}

\begin{table}[]

    \centering
    \begin{tabular}{c|cccccccc}
    Individual & Arson & Burglary & Counterfeit. & Forgery & Homicide & Kidnapp. & Mugging & Rec. st. goods \\ \hline
1&4&2&2&2&4&3&3&1 \\
2&4&2&2&2&4&4&3&1\\
3&3&2&2&2&4&3&3&1\\
4&4&3&2&2&4&4&4&3\\
5&4&3&2&2&4&4&3&2\\
6&4&3&3&2&4&4&3&2\\
7&4&1&2&2&4&4&2&1\\
8&4&4&2&2&4&4&3&2\\
9&3&2&1&2&4&4&3&1\\
10&4&3&3&3&4&4&3&2\\
11&4&2&3&3&4&4&4&1\\
12&4&4&3&3&4&4&4&2\\
13&4&3&3&2&4&4&3&1\\
14&4&2&2&2&4&3&3&1\\
15&4&2&1&1&4&4&2&1\\
16&3&2&2&2&4&3&3&1\\
17&3&2&2&2&4&4&3&2 \\
\end{tabular}
 \caption{Nishisato's 1994 seriousness of crimes rating data}
    \label{tab:DataTable}
\end{table}
The two dimensionsal dual scaling solution for the objects (crimes) can be found in Figure \ref{fig:FigDS_crimes}. In accordance with the optimal scaling formulations of \cite{Guttman1946} and \cite{Nishisato1978} the scaling values are in so-called standard coordinates. The two-dimensional solution, which is heavily dominated by the first dimension, accounts for $89$ \% of the variance. 
Correspondence analysis of the ratings results in Figure \ref{fig:FigCA_crimes}, where, in accordance with \cite[][Exhibit 23.2, p.180]{Greenacre2017}, the coordinates for the doubled objects are in principal coordinates, and, for each crime, we connected the points corresponding to the lower and upper ends of the scale, by axes. The CA solution accounts for $64$ \% of the variation. 

A one-to-one comparison of these two solutions for the objects is complicated due to the doubling of object points in the correspondence analysis solution. Moreover, we used standard coordinates in the dual scaling analysis, and principal coordinates for the correspondence analysis results. Still, comparing Figures \ref{fig:FigDS_crimes} and \ref{fig:FigCA_crimes} immediately does show a better separation of objects (crimes) in the dual scaling approach. In Figure~\ref{fig:FigDS_crimes}, we see that the crimes ``Counterfeiting" and ``Forgery", which are somewhat similar in nature, are indeed perceived as more similar by the respondents.  On the other hand, the perceptions of ``Mugging", ``Burglary" and ``Receiving stolen goods", as indicated by the ratings, differ substantially. Note that the first dimension in this analysis is rather dominant. Moreover, this dimension appears to describe mostly the perceived seriousness of crimes from more ``serious" (Arson and Kidnapping) on the left, to less ``serious" (Receiving stolen goods) on the right. 

In Figure \ref{fig:FigCA_crimes}, we see that the correspondence analysis approach (CAr) visualises that the ratings of ``Arson" and ``Kidnapping" are correlated. Furthermore, the ratings of these two crimes appear to be mostly uncorrelated to the ratings for ``Forgery", ``Mugging", ``Burglary" and ``Receiving stolen goods". % ratings also appear correlated, and slightly negatively correlated with the ratings for Arson and Kidnapping.
Note that the endpoints of the colored lines correspond to the end points of the scale. That is, the `-' points correspond to the lowest rating and the `+' points to the highest ratings. As the origin in a CA plot corresponds to average profiles, we can infer the approximate mean ratings for objects directly from the plot. For example, we see that both ``Kidnapping" and ``Arson" are rated as ``extremely serious" far more often than average. Similarly, ``Receiving stolen goods" tends to receive a lower (less serious) rating more often than not. For ``Burglary", the results are more varied and the average rating appears to be close to the middle of the rating scale. 

\begin{figure}[ht]
    %\centering
    \begin{center}
\includegraphics[scale=.3]{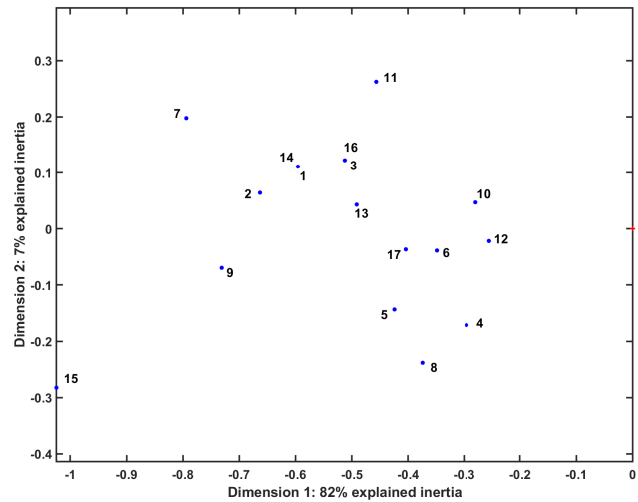}
    \caption{Dual scaling of crime rating data. Scores for individuals in principal coordinates}
    \label{fig:FigDS_crimes_subjects}
\end{center}
    
\end{figure}

\begin{figure}[ht]
    %\centering
    \begin{center}
\includegraphics[scale=.3]{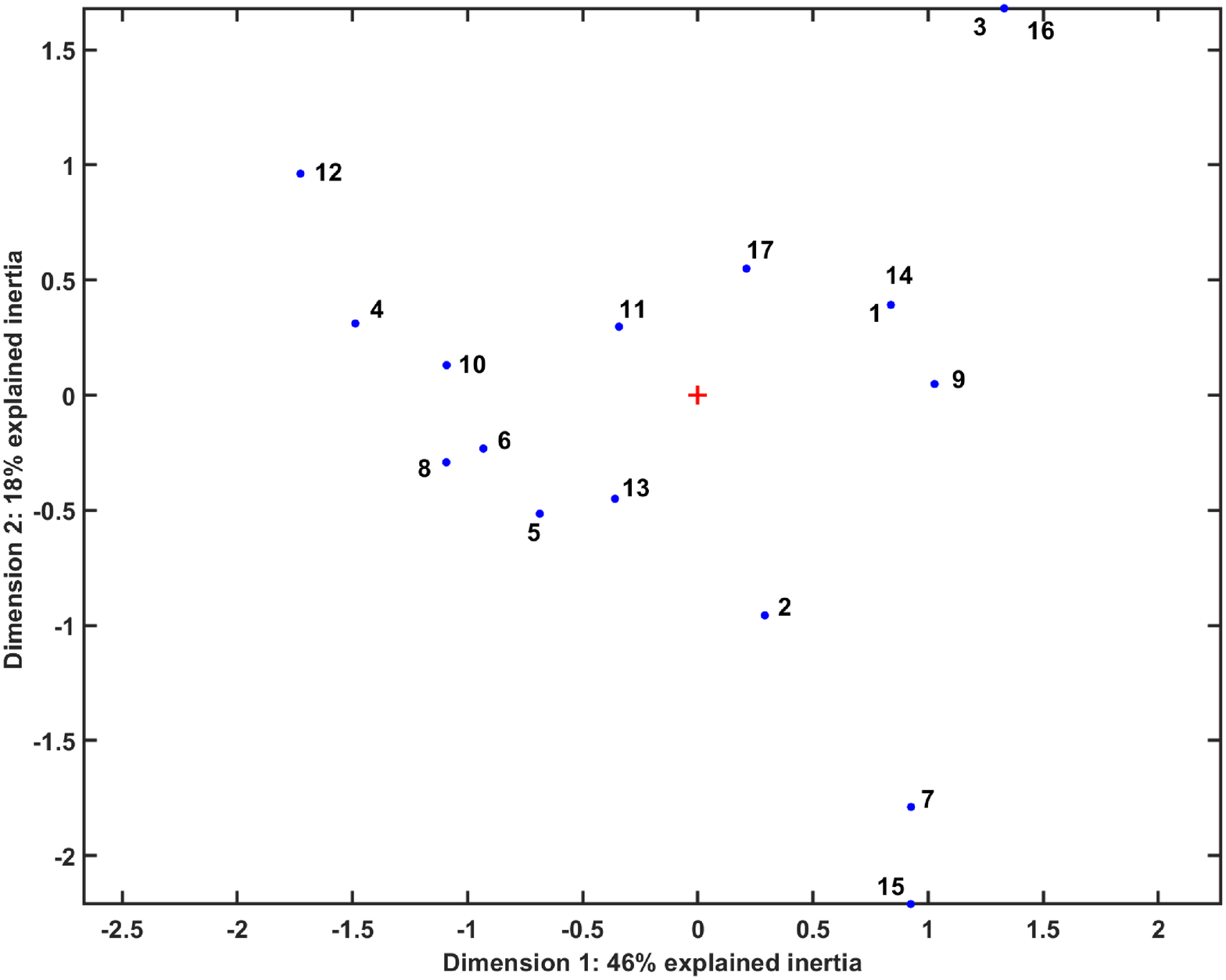}
    \caption{Correspondence analysis of crime rating data. Optimal scaling values for individuals in standard coordinates}
    \label{fig:FigCA_crimes_subjects}
\end{center}
    
\end{figure}

Figures \ref{fig:FigDS_crimes_subjects}, for DS3, and \ref{fig:FigCA_crimes_subjects}, for CAr, give, for both analyses, the corresponding solutions for the individuals. Hence, for the dual scaling solution, the scores for the individuals are in principal coordinates whereas for the CA solution they are in standard coordinates. In addition, the doubled set of ``individual" scores for the dual scaling solution is ignored as these are simply the same coordinates mirrored in the origin. 

Recall that the correspondence analysis solution gives optimal scaling values for the individuals. Hence, coordinates are determined in such a way that differences in the indicated rating patterns between individuals is optimally depicted. Superficially comparing Figures \ref{fig:FigDS_crimes_subjects} and \ref{fig:FigCA_crimes_subjects} may not immediately expose this. However, note that for the dual scaling solution, depicted by Figure \ref{fig:FigDS_crimes_subjects}, the points are not spread out along both dimensions. Instead, they are all concentrated on the negative side of the first dimension. 

\begin{figure}[ht]
    %\centering
    \begin{center}
\includegraphics[scale=.3]{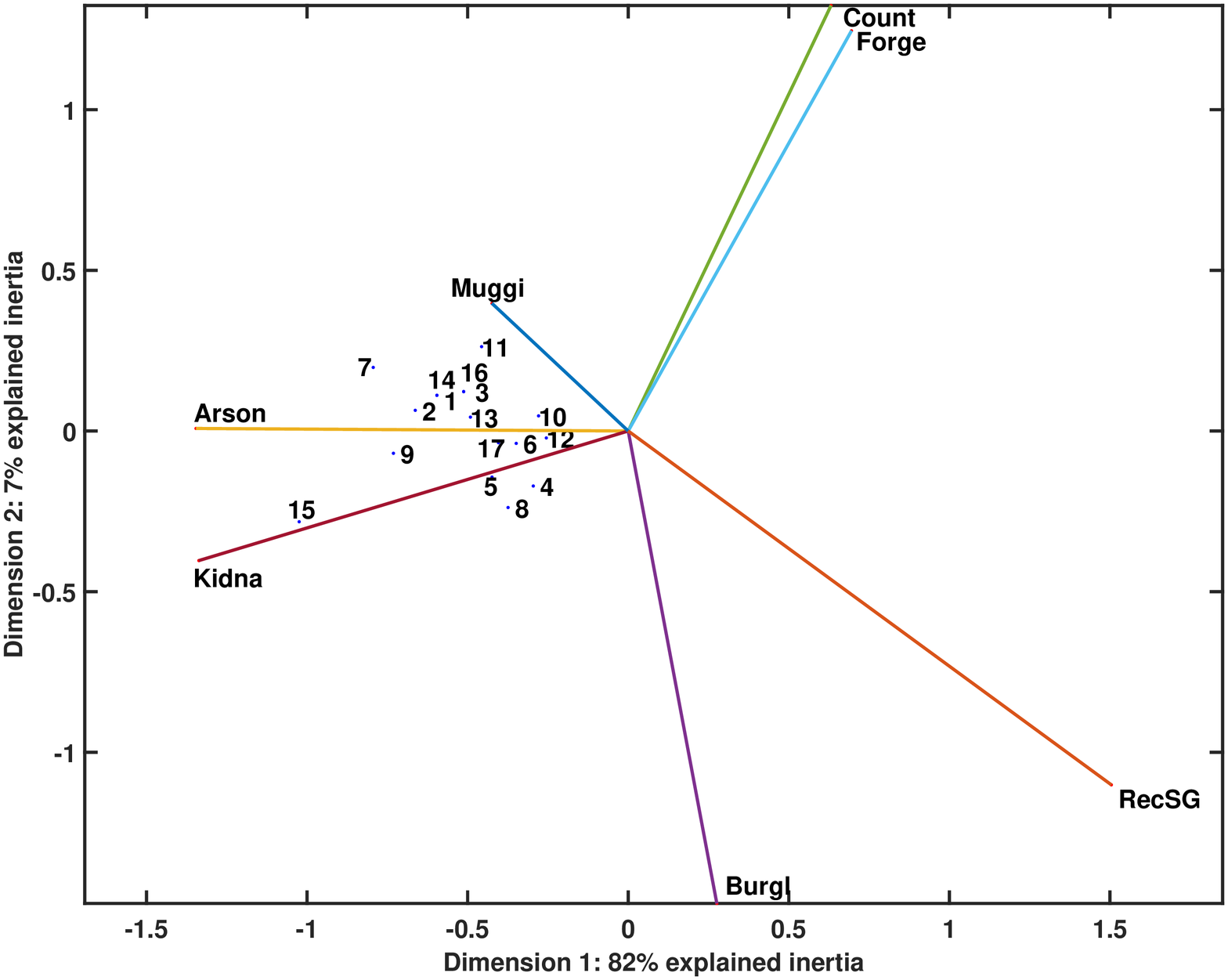}
    \caption{Dual scaling biplot of crime rating data. Optimal scaling values for objects (crimes) in standard coordinates}
    \label{fig:FigDS_biplot}
\end{center}
    
\end{figure}

\begin{figure}[ht]
    %\centering
    \begin{center}
\includegraphics[scale=.3]{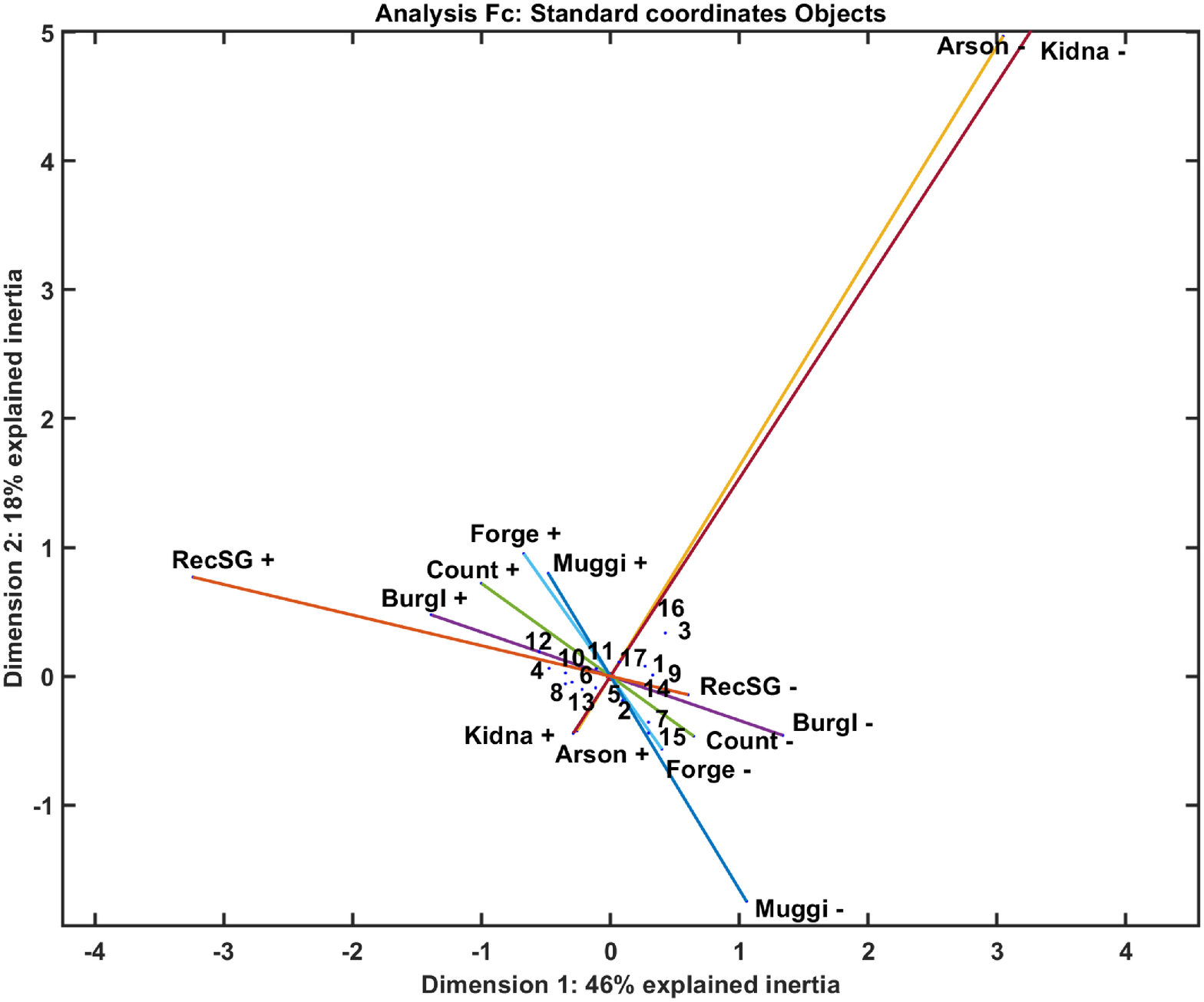}
    \caption{Correspondence analysis biplot of crime rating data. Optimal scaling values for individuals in principal coordinates}
    \label{fig:FigCA_biplot}
\end{center}

\end{figure}

To better appreciate the differences in the solutions, Figures \ref{fig:FigDS_biplot} and \ref{fig:FigCA_biplot} give biplots for both methods. That is, joint plots for rows and columns where projections of one set of points on the directions of other points (obtained, for example, by drawing axes from the origin through the points), can be used to reconstruct the values in the original data table. See \cite{Greenacre1993} for more details. Note that, in both joint plots objects (crimes) are in standard, and individuals are in principal coordinates. 

Interpreting the relative positions of the individuals in Figure \ref{fig:FigDS_biplot} is not so easy. For these data, differences are small and most individuals give high ratings to all crimes. That is, they tend to find all crimes to be \emph{serious}. How individual $15$ differentiates from the others, as its location in Figures \ref{fig:FigDS_crimes_subjects} and \ref{fig:FigDS_biplot} suggests, is not clear from the plot. 

The optimal scaling positions of individuals in Figure \ref{fig:FigCA_crimes_subjects} appear better separated. Moreover, the interpretation of the differences in the locations of the individuals is more straightforward. For example, individuals $15$ and $7$ are separated from the other points. In the biplot of Figure \ref{fig:FigCA_biplot}, we see that this may be explained by both individuals giving relatively low ratings (that is, a lower rating than average) for ``Mugging". Indeed, these two individuals are the only ones that assign a rating $2$ to these two crimes. All others give higher ratings. In a similar way, differences in positions of other ``outlying" points (e.g., $3, 16, 12, 4)$ can be explained by observing in what sense the corresponding rating profiles differ from the average rating profiles. For the equivalent ratings of individuals $3$ and $16$, we see that they differ from all other individuals with respect to their rating for ``Arson" and ``Kidnapping". As can be verified from the data table, they gave a rating of $3$ to both of these crimes whereas all others either gave rating $4$ to at least one of these crimes. 

\section{Conclusion} \label{SectFinal}

In this paper, that has been inspired by the works of Nishisato, we introduced a dual scaling of rating approach. We showed how this method relates to the correspondence analysis of rating data and that the fundamental difference between these two variants can be attributed to a difference in pre-processing of the data. In particular, the dual scaling of rating data can be described as dual scaling of a row-wise doubled matrix whereas correspondence analysis amounts to the analysis of a column-wise doubled matrix. 

The dual scaling framework that has been laid out by Nishisato throughout his career offers tools to better understand the resulting differences. That is, whereas the dual scaling of rating data yields (and in fact, was defined to do so) optimal scaling values for the objects, the correspondence analysis of rating data yields optimal scaling values for the individuals. Given these rather fundamental differences, saying that one approach is better than the other, does not make much sense. A choice between these two variants depends on the research goals. If the goal is to find scale values (or: a representation) that best separates the objects according to the observed ratings, the dual scaling of ratings (that is: the analysis of the row-wise double matrix $\mathbf{F}_{r}$) is appropriate. On the other hand, to better distinguish individuals according to their ratings, correspondence analysis of ratings (that is: the analysis of the column-wise doubled matrix $\mathbf{F}_{c}$) is the better alternative. 

\bibliography{dsratings_ref.bib}
\bibliographystyle{apalike}

\end{document}